\documentclass[sigconf,natbib=true]{acmart}

\setcopyright{none}
\renewcommand\footnotetextcopyrightpermission[1]{}
\usepackage{multirow}
\usepackage{booktabs}
\usepackage{enumitem}
\usepackage{xurl}
\usepackage{tabularx}

\AtBeginDocument{
  }

\acmConference[arXiv Preprint]{arXiv Preprint}{}{}

\begin{document}

\title{RWGBench: Evaluating Scholarly Positioning in Related Work Generation}

\author{Anzhe Xie}
\email{xaz25@mails.tsinghua.edu.cn}
\affiliation{%
  \institution{Tsinghua University}
  \city{Beijing}
  \country{China}
}

\author{Weihang Su}
\email{swh22@mails.tsinghua.edu.cn}
\affiliation{%
  \institution{Tsinghua University}
  \city{Beijing}
  \country{China}
}

\author{Jiaxin Mao}
\email{maojiaxin@gmail.com}
\affiliation{%
  \institution{Renmin University of China}
  \city{Beijing}
  \country{China}
}

\author{Yiqun Liu}
\email{yiqunliu@tsinghua.edu.cn}
\affiliation{%
  \institution{Tsinghua University}
  \city{Beijing}
  \country{China}
}

\author{Shaoping Ma}
\email{msp@tsinghua.edu.cn}
\affiliation{%
  \institution{Tsinghua University}
  \city{Beijing}
  \country{China}
}

\author{Qingyao Ai}
\authornote{Corresponding author.}
\email{aiqingyao@gmail.com}
\affiliation{%
  \institution{Tsinghua University}
  \city{Beijing}
  \country{China}
}
\renewcommand{\shortauthors}{Xie et al.}

\begin{abstract}
Large language models have shown strong fluency in scientific writing, yet the evaluation of related work generation (RWG) remains limited. 
Existing RWG evaluations largely inherit summarization-oriented metrics, using lexical or semantic similarity to reference sections as proxies for quality. However, related work writing is fundamentally a citation-level scholarly positioning task: it requires selecting, organizing, and framing prior work to clarify how a target paper relates to, differs from, and contributes beyond existing research.
As a result, models may generate coherent and semantically relevant text while exhibiting academically critical failures, such as inappropriate citation selection or misplaced references, that conventional metrics do not capture.
To this end, we introduce \textbf{RWGBench}, a benchmark that evaluates RWG from the perspective of citation decision-making rather than text similarity. 
RWGBench is constructed from a large-scale collection of 40,108 computer science papers and a retrieval corpus of 1.09 million documents, with a peer-reviewed test set comprising 100 papers accepted at ICLR, NeurIPS, or ICML and their corresponding author-written related work sections.
We propose a multi-dimensional evaluation framework that assesses citation selection, contextual appropriateness, organization, and citation framing.
Experiments across representative and frontier generation settings reveal systematic limitations that are obscured by standard evaluations, separating retrieval bottlenecks from generation-level positioning failures. A blinded human study provides supplementary support for the proposed diagnostic metrics.
RWGBench offers a citation-centric testbed for developing and evaluating related work generation systems that are better aligned with scholarly writing practices.
\end{abstract}

\begin{CCSXML}
<ccs2012>
<concept>
<concept_id>10002951.10003317.10003347.10003350</concept_id>
<concept_desc>Information systems~Evaluation of retrieval results</concept_desc>
<concept_significance>500</concept_significance>
</concept>
<concept>
<concept_id>10002951.10003260.10003309.10003315</concept_id>
<concept_desc>Information systems~Summarization</concept_desc>
<concept_significance>500</concept_significance>
</concept>
<concept>
<concept_id>10010147.10010178.10010187</concept_id>
<concept_desc>Computing methodologies~Natural language generation</concept_desc>
<concept_significance>300</concept_significance>
</concept>
</ccs2012>
\end{CCSXML}

\ccsdesc[500]{Information systems~Evaluation of retrieval results}
\ccsdesc[500]{Information systems~Summarization}
\ccsdesc[300]{Computing methodologies~Natural language generation}

\keywords{Related work generation, citation recommendation, scholarly positioning, benchmark, LLM}

\maketitle
\footnotetext[1]{Anonymous repository: \url{https://anonymous.4open.science/r/RWGBench-7642}}

\section{Introduction}

The rapid growth of scientific literature has made it increasingly difficult for researchers to situate new work within a rapidly evolving research landscape. In computer science alone, more than 200,000 articles are published annually~\cite{khalid2025comprehensive,bornmann2021growth}, substantially increasing the burden of identifying relevant prior work, explaining how a study differs from existing research, and articulating its contribution. This challenge has motivated growing interest in automated related work generation (RWG), in which large language models~\cite{gpt4,qwen3,llama} assist authors in drafting related work sections conditioned on a target paper.

Despite this growing interest, related work generation is not merely a special case of generic summarization or survey generation.
While survey generation typically seeks to synthesize a broad body of topic-level literature with relatively comprehensive coverage~\cite{autosurvey,surge,gao2024llm}, RWG is anchored to a specific target paper and serves a more targeted rhetorical function: situating that paper within prior research. This process, often characterized as \textit{scholarly positioning}~\cite{ZHANG2022101076,PETRIC2007238,Swales2004}, requires selecting, organizing, and framing prior work so as to clarify the target paper's relevance, differentiation from existing studies, and contribution to the field.
A high-quality related work section, therefore, does more than describe previous studies: it decides which works should be cited, how they should be grouped, and how their relationship to the target paper should be expressed. These interdependent citation decisions jointly shape how the target paper is perceived in the literature, making RWG a structured citation-decision task rather than a purely surface-level text-generation task~\cite{teufel2006}.

Existing evaluations for RWG remain poorly aligned with this distinctive function. Most current metrics are inherited from summarization and long-form generation~\cite{novikova2017we,gehrmann2022repairing}, emphasizing lexical overlap, semantic similarity, or general fluency with respect to human-written reference sections. Such metrics can reward outputs that resemble the reference text while failing to detect academically critical errors in citation behavior. 
For example, a generated related work section may achieve high textual similarity while selecting poorly matched references, over-relying on a narrow subset of citations, omitting central prior work, or placing citations in contexts that misrepresent their relevance. 
Meanwhile, existing benchmarks mainly focus on sentence-level citation generation or survey-style literature synthesis~\cite{cohan2019structural,xing2020automatic,surge,li2024related,deepscholar2025,docekal2024oarelatedwork}, and therefore provide limited support for evaluating document-level citation selection, citation placement, organizational structure, and discourse-level positioning. As a result, current evaluations offer only a partial picture of RWG quality and provide limited diagnostic insight into how models fail when positioning a target paper within prior research.

To address this gap, we introduce \textbf{RWGBench}, a benchmark for evaluating RWG as a \textit{citation decision-making process}. 
RWGBench is constructed from 40,108 computer science papers and a retrieval corpus containing 1.09 million candidate documents. 
We further curate a high-fidelity test set of 100 papers accepted at peer-reviewed venues, where extracted related-work citation entries have a 97.9\% average per-paper match rate to the retrieval corpus. Building on this resource, we propose a multidimensional evaluation framework that goes beyond surface-level text similarity. The framework assesses the core citation decisions that underlie scholarly positioning, including citation selection, placement, organization, and rhetorical framing. By evaluating both generated text and the citation decisions that support it, RWGBench enables a more faithful and diagnostic assessment of modern LLMs for related work generation.\footnote{Code, metadata, and anonymized documentation will be released upon acceptance.}

In summary, our contributions are as follows.
\begin{enumerate}[leftmargin=*]
    \item We reformulate RWG as a scholarly positioning task centered on citation decision-making.
    \item We construct RWGBench, a realistic benchmark with curated evaluation targets and a large-scale retrieval corpus.
    \item We propose a behavior-oriented evaluation framework enabling fine-grained diagnosis of RWG failure modes.
    \item We provide an empirical analysis on RWGBench that exposes critical limitations of representative RWG systems, including a frontier GPT-5.5 generator suite.
\end{enumerate}

\section{Task Definition and Problem Formalization}
\label{sec:task}

This section defines related work generation as a document-level task conditioned on a target paper. We then highlight its central challenge: scholarly positioning, in which citation selection, placement, and framing jointly determine how the target paper is situated within the prior literature.

\subsection{Task Formulation}

We formulate RWG as a document-level generation task conditioned on a target paper $P$, which consists of a title $T$, an abstract $A$, and optionally an introduction $I$, along with a large corpus of candidate papers $\mathcal{D} = \{d_1, d_2, \ldots, d_n\}$. 
The objective is to retrieve a relevant subset of papers $C$ from $\mathcal{D}$, then generate a related work section $R$ that organizes them into a coherent narrative structure, and positions $P$ effectively within its broader research context.
In practice, RWG is typically implemented as a two-stage process:
\begin{enumerate}[leftmargin=*]
    \item \textbf{Retrieval:} $\text{Retrieve}(P, \mathcal{D}) \rightarrow C$, identifying candidate papers to cite;
    \item \textbf{Generation:} $\text{Generate}(P, C) \rightarrow R$, synthesizing the related work text with citations.
\end{enumerate}

While this formulation resembles retrieval-augmented generation, the core challenge of RWG lies not in text synthesis alone, but in the quality of the citation decisions embedded in both stages.

\subsection{Scholarly Positioning}

Unlike generic text generation or summarization tasks, related work generation involves multi-faceted scholarly decision-making, requiring authors to jointly reason over three interdependent dimensions~\cite{ZHANG2022101076,PETRIC2007238}.
\begin{enumerate}[leftmargin=*]

\item \textbf{Citation Selection.}
Given a set of topically related papers, authors must determine which works to cite based on their relevance to the target paper's context, limitations, and contributions.

\item \textbf{Strategic Placement.}
Citation placement reflects discourse structure and guides readers through the logical development of the paper, shaping how evidence and claims are interpreted~\cite{PETRIC2007238}.

\item  \textbf{Rhetorical Framing.}
Authors must further decide how cited works are framed, for example, whether they build upon or contrast with prior approaches, influencing how relationships between studies are understood.
\end{enumerate}

Consequently, a valid RWG system must go beyond surface-level text generation to model these underlying decisions, ensuring that the generated narrative effectively clarifies the target paper's relevance and differentiation~\cite{Swales2004}.

\section{Dataset Construction}
\label{sec:data}

This section describes how RWGBench is constructed. We detail the collection of related work sections, the preparation of the retrieval corpus, the curation of the test set, the resulting dataset statistics, and licensing considerations.

\subsection{Related Work Collection}

We construct RWGBench by collecting computer science papers from arXiv. We first use the official arXiv API to obtain papers with their title, abstract, and full HTML source. We focus on papers published between 2020 and 2025 to ensure topical relevance and methodological currency.
Structured content is parsed directly from publicly available HTML documents, enabling scalable and reliable extraction. We identify sections corresponding to related work discussions, including \textit{Related Work}, \textit{Literature Review}, and semantically equivalent variants.
This process yields 40,108 computer science papers with successfully extracted metadata, citation lists, and related work sections, forming the foundation of our benchmark.

\subsection{Corpus Preparation}

To support realistic citation retrieval and grounding, we construct a large-scale retrieval corpus. We begin with the publicly released SurGE corpus, which provides broad coverage of scientific documents.
We align citations extracted from related work sections with entries in the SurGE corpus. For cited papers missing from the corpus, we further query the arXiv API and merge successfully retrieved documents after normalization. This hybrid strategy combines large-scale coverage with citation alignment, producing a corpus suitable for end-to-end RWG evaluation.
The final retrieval corpus contains 1,091,394 papers.

\subsection{Test Set Construction}

A high-quality test set is essential for reliable diagnosis of citation behavior. We therefore adopt a multi-stage construction process that balances extraction reliability, citation coverage, and source-paper quality.
From the full collection, we first filter papers based on extraction reliability and citation statistics. We retain papers with citation extraction rates of at least 95\% and moderate citation counts between 10 and 50, excluding cases with sparse or excessive references. This step yields 2,324 high-quality candidates.
We then match these candidates to accepted ICLR, NeurIPS, and ICML papers identified via OpenReview metadata by normalized title. This yields a peer-reviewed evaluation split with author-written related work sections. We select 100 exact title matches, covering ICLR (42 papers), NeurIPS (34 papers), and ICML (24 papers), with publication years from 2023 to 2025. This shared split is evaluated across 52 system configurations, producing 5,200 generated sections for automatic comparison.
Each test paper has at least 95\% of its extracted related-work citation entries matched to the retrieval corpus, with an average per-paper match rate of 97.9\%. Evaluation uses these matched citation identifiers as the reference citation set, and every identifier in this set is present in the retrieval corpus. This keeps missing-corpus artifacts small while making the citation-mapping boundary explicit.
\subsection{Statistics and Analysis}

\begin{table}[t]
\centering
\caption{Dataset statistics of RWGBench, including the full dataset and the test set. Matched citation coverage is the per-paper average of matched citation entries divided by all citation entries extracted from each related work section.}
\label{tab:data_stats}
\begin{tabular}{@{}lcc@{}}
\toprule
\textbf{Statistic} & \textbf{Full Dataset} & \textbf{Test Set} \\
\midrule
Papers & 40,108 & 100 \\
Avg. Extracted Citations & 26.0 & 25.3 \\
Avg. Matched Citations & 14.7 & 24.7 \\
Avg. RW Length (words) & 673 & 687 \\
Avg. Abstract Length (words) & 192 & 200 \\
Matched Citation Coverage & 58.2\% & 97.9\% \\
\midrule
% \multicolumn{3}{l}{\textit{Retrieval Corpus}} \\
Corpus Size & \multicolumn{2}{c}{1,091,394} \\
Corpus Avg. Abstract Length (words) & \multicolumn{2}{c}{164} \\
\bottomrule
\end{tabular}
\end{table}

Table~\ref{tab:data_stats} summarizes the key statistics of RWGBench. The benchmark comprises 40,108 target papers, with a held-out peer-reviewed test set of 100 papers, paired with a large-scale retrieval corpus containing 1,091,394 documents. The full collection supports benchmark construction and corpus analysis, while citation-based system evaluation is reported on the high-coverage peer-reviewed test set using matched citation identifiers.
The related work sections exhibit moderate length and citation density, with an average of 673 words and 14.7 matched citation identifiers per paper in the full dataset. Test papers are comparatively more citation-intensive, with 25.3 extracted citation entries and 24.7 matched citation identifiers per paper on average, which increases the difficulty of citation selection and placement.
The retrieval corpus covers an average of 58.2\% of extracted related-work citation entries per paper in the full dataset, as some cited papers originate from venues not indexed by arXiv. For the peer-reviewed test set, this matched citation coverage rises to 97.9\%. All matched citation identifiers used as gold labels in evaluation are present in the corpus, allowing Oracle and retrieval-based experiments to isolate generation, selection, and framing errors over a clearly defined reference set.

\subsection{Ethical Considerations and Licensing}

All papers, including the 100 peer-reviewed test papers, are publicly accessible through arXiv and/or OpenReview. We parse structured content from public HTML sources, including titles, abstracts, introductions, and related work sections; no full-text PDFs are redistributed. Because arXiv papers are released under heterogeneous licenses, the public release separates code from document-derived content. It will provide paper identifiers, citation identifiers, metadata, and extraction scripts, while redistributing text only when the corresponding license permits it. All included papers are public scholarly documents and contain no personally identifiable information beyond standard author metadata.
RWGBench is designed to evaluate and diagnose LLM behavior in scholarly writing tasks. We view rigorous benchmarking as essential for understanding where LLM assistance is appropriate and where it falls short, particularly in tasks as consequential as scholarly positioning.

\section{Evaluation Framework}
\label{sec:eval}

Evaluating related work generation requires assessing not only surface-level text quality, but also the scholarly decisions underlying citation use. To this end, we design a multi-dimensional evaluation framework that targets different aspects of \textit{scholarly positioning}.
For reporting, we organize the metrics into four groups: Citation Quality (CQ), Citation Organization (CO), Citation Framing (CF), and Text Quality (TQ). CQ covers both citation-set selection and local citation appropriateness; CO and CF capture section-level placement and rhetorical use; TQ captures surface-level fluency and coherence.

\subsection{Citation Quality}

Citation quality evaluates whether a model selects papers that genuinely support its academic argumentation, rather than relying on superficial topical similarity.

\subsubsection{Citation Precision, Recall, and Soft Recall}

We first compare the set of generated citations $\mathcal{C}_g$ against the canonical reference citation set $\mathcal{C}_r$ used by the original authors. Citation precision and recall are defined as:

\begin{align}
P_{\text{cite}} = \frac{|\mathcal{C}_g \cap \mathcal{C}_r|}{|\mathcal{C}_g|}\quad,\quad 
R_{\text{cite}} = \frac{|\mathcal{C}_g \cap \mathcal{C}_r|}{|\mathcal{C}_r|}.
\end{align}

\noindent Here, $\mathcal{C}_g$ denotes the set of citations generated by the model, and $\mathcal{C}_r$ denotes the reference citation set from the author-written related work section. $P_{\text{cite}}$ and $R_{\text{cite}}$ measure alignment between generated citations and canonical references used by human authors. In the experiments, generated inline citation markers are resolved deterministically through the ordered candidate list supplied to the model, so citation evaluation does not rely on post-hoc fuzzy title matching.

Related work writing can contain multiple defensible citation choices, since different prior papers may play similar topical roles. We therefore additionally report a soft citation recall that gives partial credit when a generated citation is topically compatible with a reference citation. Let $s_{\mathrm{NLI}}(g,r)$ denote an NLI-based topic-compatibility score between generated citation $g$ and reference citation $r$. The generated citation's title and abstract are used as the premise, and the reference citation's title and abstract are used as the hypothesis. We map NLI probabilities to a graded score as $P(\textsc{Entailment}) + 0.5P(\textsc{Neutral})$, assigning no credit to contradiction. Soft recall is computed as:
\begin{align}
R_{\text{soft}} =
\frac{1}{|\mathcal{C}_r|}
\sum_{r \in \mathcal{C}_r}
\max_{g \in \mathcal{C}_g} s_{\mathrm{NLI}}(g,r).
\end{align}
\noindent This metric keeps the author citation set as the canonical reference while recognizing topically plausible alternatives.

\subsubsection{Claim-Support Consistency (CSC)}

Correct citation further requires that cited papers support the specific claims in which they appear. We therefore adopt an NLI-based verification approach inspired by fact verification methods~\cite{thorne2018fever}.

For each generated citation $c_i \in \mathcal{C}_g$, we construct:
\begin{itemize}[leftmargin=*]
    \item Premise $p_i$: the abstract of the cited paper $c_i$,
    \item Hypothesis $h_i$: the sentence in which $c_i$ is cited.
\end{itemize}

We apply a DeBERTa-v3 NLI model~\cite{he2021deberta} to predict the natural language inference relation between $(p_i, h_i)$. The predicted labels entailment, neutral, and contradiction are mapped to scores of $1.0$, $0.5$, and $0.0$, respectively. The citation context score is computed as:

\begin{align}
CSC = \frac{1}{|\mathcal{C}_g|} \sum_{c_i \in \mathcal{C}_g} s_i,
\end{align}

\noindent where $s_i$ denotes the NLI-based support score for citation $c_i$. $CSC$ evaluates whether selected citations appropriately support the claims in which they appear, reflecting the validity of citation selection at the local argumentative level.

\subsubsection{Topical Relevance to Target Work (TRT)}

While $CSC$ focuses on sentence-level support, we further report topical relevance at the document level as a diagnostic descriptor. Let $P$ denote the target paper and $c_i \in \mathcal{C}_g$ a generated citation. We compute semantic similarity between abstracts using SciBERT embeddings~\cite{beltagy2019scibert}. The citation topic score is defined as:

\begin{align}
TRT = \frac{1}{|\mathcal{C}_g|} 
\sum_{c_i \in \mathcal{C}_g} 
\mathrm{sim}\!\left( \mathbf{a}_P,\ \mathbf{a}_{c_i} \right),
\end{align}

\noindent where $\mathbf{a}_P$ and $\mathbf{a}_{c_i}$ denote the abstract embeddings of the target paper and cited paper $c_i$, respectively, and $\mathrm{sim}(\cdot,\cdot)$ denotes cosine similarity. $TRT$ captures whether selected citations are topically aligned with the target paper at the document level. We interpret it as a necessary but insufficient signal for citation quality, since topical proximity alone does not determine whether a paper should be cited.

\subsection{Citation Organization}

This dimension evaluates how models allocate and distribute citations across the related work section, reflecting strategic placement and emphasis decisions essential for scholarly positioning.
\subsubsection{Citation Density (CD)}

We measure citation density to detect under-citing or over-citing behaviors. Let $\mathcal{C}_g$ denote the set of generated citations and $R_g$ the generated related work section, consisting of $N_s$ sentences. Citation density is defined as:
\begin{align}
\mathrm{CD} = \frac{|\mathcal{C}_g|}{N_s}.
\end{align}
\noindent Citation density evaluates whether models allocate citations at a rate comparable to human-authored related work sections. Extreme values indicate positioning failures, such as insufficient grounding or excessive citation clustering that obscures argumentative focus.

\subsubsection{Citation Distribution Balance (CDB)}

To assess how citations are distributed across the section, we compute the proportion of cited papers that appear exactly once. For each citation $c \in \mathcal{C}_g$, let $f(c)$ denote its frequency of occurrence in $R_g$. The unique citation ratio is defined as:

\begin{align}
\mathrm{CDB} = 
\frac{|\{ c \in \mathcal{C}_g \mid f(c) = 1 \}|}{|\mathcal{C}_g|}.
\end{align}

\noindent CDB reflects the balance between citation diversity and depth. Extremely low values indicate repetitive reliance on a small set of papers, while overly high values may signal shallow engagement with prior work.

\subsection{Citation Framing and Structural Coherence}

This dimension evaluates whether generated related work sections exhibit global rhetorical structures consistent with scholarly positioning conventions~\cite{PETRIC2007238,ZHANG2022101076}, such as contextualization, gap identification, and contribution positioning.

\subsubsection{Positioning Structure Similarity (PSS)}

Related work sections are organized sequences of content states rather than unordered bags of sentences. Following content-model approaches to discourse organization~\cite{barzilay-lee-2004-catching} and LCS-based sequence evaluation~\cite{lin2004rouge}, we segment generated and reference related work sections into ordered content blocks. Each block is embedded with SciBERT~\cite{beltagy2019scibert}, after which we compute a soft longest-common-subsequence alignment. Let $u^g_{1:m}$ and $u^r_{1:n}$ denote generated and reference blocks, and let $a_{ij}=\max(0,\cos(u^g_j,u^r_i))$ be the non-negative SciBERT cosine similarity between $u^g_j$ and $u^r_i$. The soft LCS score $L$ is computed by dynamic programming:
\begin{align}
L_{ij} = \max\{L_{i-1,j}, L_{i,j-1}, L_{i-1,j-1}+a_{ij}\}.
\end{align}
We define:
\begin{align}
\mathrm{PSS} = \frac{2L_{mn}}{m+n}.
\end{align}
\noindent PSS evaluates whether generated sections cover similar positioning content in a comparable order, penalizing missing, extra, or reordered content blocks. Since related work sections can be organized in multiple defensible ways, PSS is used as a reference-relative structural diagnostic rather than an absolute judgment that the author order is uniquely correct.

\subsubsection{Citation Frame Alignment (CFA)}

PSS captures broad discourse structure, but it does not directly measure how citations are used as scholarly positioning devices. We therefore introduce Citation Frame Alignment (CFA), which compares the citation-function profile of generated and reference sections. Building on citation-intent and citation-function studies~\cite{multicite2021,cohan2019structural,jurgens2018citation,PETRIC2007238,teufel2006}, we adopt the seven MultiCite labels: motivation, background, uses, extends, similarities, differences, and future work. These labels capture how citations contextualize a problem, justify choices, connect methods, contrast novelty, and point to future directions.

For each valid citation-bearing context, a MultiCite-trained classifier predicts one or more citation frames. We aggregate frame counts over the generated and reference sections as $\mathbf{y}_g$ and $\mathbf{y}_r$. For each active frame $k$, alignment is measured by count overlap:
\begin{align}
\mathrm{F}_k = \frac{2\min(y_{g,k},y_{r,k})}{y_{g,k}+y_{r,k}},
\quad
\mathrm{CFA}=\frac{1}{|\mathcal{A}|}\sum_{k\in\mathcal{A}}\mathrm{F}_k,
\end{align}
where $\mathcal{A}=\{k \mid y_{g,k}+y_{r,k}>0\}$. For generated text, we count only citation contexts whose citation marker maps to a retrieved or generated paper. CFA complements citation selection metrics by testing whether valid citations play similar rhetorical roles in the generated and author-written sections. Like PSS, CFA is intended as a comparative diagnostic of framing behavior, not as a claim that the reference distribution is the only acceptable rhetorical profile.

\subsection{Text Quality}

While scholarly positioning is the primary focus of our evaluation, we additionally assess surface-level text quality to disentangle decision-making failures from purely linguistic deficiencies.

\begin{itemize}[leftmargin=*]

\item \textbf{ROUGE-L}~\cite{lin2004rouge} measures longest common subsequence overlap with the reference, capturing lexical similarity.

\item \textbf{BERTScore}~\cite{zhang2019bertscore} evaluates semantic similarity using contextual embeddings based on SciBERT, providing robustness to paraphrasing.

\item \textbf{LLM-as-Judge.} Following recent work~\cite{zheng2023judging,liu2023geval,gu2025surveyllmasajudge,shi2025deepresearchsystematicsurvey}, we employ an LLM-based evaluator to assign a single publication-readiness score from 1 to 5 for each related work section. The prompt focuses on coherence, organization, fluency, and academic writing style, while explicitly excluding citation correctness and citation coverage, which are measured separately. We report the score after linear normalization to $[0,1]$.

\end{itemize}

\noindent Collectively, these dimensions operationalize scholarly positioning as a set of observable selection, allocation, and structuring behaviors, enabling fine-grained diagnosis of RWG failures beyond surface-level fluency.

\section{Experiments and Analysis}
\label{sec:exp}

We conduct comprehensive experiments to demonstrate the diagnostic utility of RWGBench and reveal systematic challenges in current approaches.

\subsection{Implementation Details}

We evaluate Direct generation and four retrieval-augmented generation baselines with three representative LLMs (DeepSeek-V3~\cite{deepseek}, Llama-3-8B~\cite{llama}, Qwen-2.5-7B~\cite{qwen}). We additionally run the full baseline suite with GPT-5.5 under the same retrieval settings as a frontier generator check. BM25 uses $k_1=1.5$ and $b=0.75$ over 1.09M documents. Dense retrieval uses the off-the-shelf BGE-large-en-v1.5 encoder~\cite{bge} with FAISS~\cite{faiss} indexing and no task-specific retriever fine-tuning. RAG, ALCE, and TAG use the top-50 retrieved papers. Workflow retrieves multiple top-20 lists from planned queries, deduplicates them, and passes at most 50 ranked candidates to the generator.
For generation, we set the temperature to $0.7$ and max tokens to $2048$. GPT-5.5 is queried through an OpenAI-compatible API. Models receive the target paper (title, abstract, introduction) and retrieved papers (title + abstract), and are instructed to generate related work with inline citations. Each citation marker $[i]$ is mapped back to the $i$-th supplied candidate paper; invalid or out-of-range markers are not counted as resolved citations.
For evaluation, CSC and soft citation recall use a DeBERTa-v3 NLI model~\cite{he2021deberta}; PSS and BERTScore use SciBERT~\cite{beltagy2019scibert}; CFA uses a MultiCite-trained citation-function classifier~\cite{multicite2021}. For the auxiliary LLM-as-Judge text-quality score, GPT-5.5 is the default scorer, with Claude Opus and Gemini Pro used for agreement analysis.

\subsection{Baseline Systems}

We compare the following representative related work generation approaches.\footnote{The implementation of \textit{RAG} and \textit{Workflow} will be released with the benchmark.}

\begin{itemize}[leftmargin=*]

\item \textbf{RAG}: Standard retrieval-augmented generation. Models receive top-$k=50$ retrieved papers (title + abstract) and generate related work with citations.

\item \textbf{ALCE}: Adapted from the grounded citation generation framework~\cite{alce}. While the original work explores both prompt-based and fine-tuning approaches, we implement the prompt-based vanilla variant with full abstracts as numbered passages, emphasizing explicit grounding instructions without model training.

\item \textbf{TAG}: Adapted from the target-aware related work generation method~\cite{tag}. We implement a prompt-based adaptation that simulates target-awareness through keyphrase-guided organization and explicit modeling of the target paper's relationship to retrieved papers, without specialized training.

\item \textbf{Workflow}: Multi-stage generation that first performs theme-based planning, then issues multiple retrieval queries ($k=20$ per query), evaluates citation coverage, and iteratively refines the generated content (up to 2 iterations).

\item \textbf{Direct}: Directly prompt the LLM to generate the related work with no extra information.

\item \textbf{Oracle}: Uses ground-truth citations directly, bypassing retrieval to isolate generation-level failures and establish an upper bound.

\end{itemize}

\subsection{Main Results}

\begin{table*}[t]
\centering
\caption{Experimental results across all models, baselines, and retrievers. All metrics are averaged over 100 peer-reviewed test papers.}
\label{tab:main_results}

\setlength{\tabcolsep}{6.2pt}
\renewcommand{\arraystretch}{0.82}
\setlength{\aboverulesep}{0.28ex}
\setlength{\belowrulesep}{0.28ex}

\resizebox{\textwidth}{!}{%
\begin{tabular}{lll|ccccc|cc|cc|ccc}
\toprule

 &  &  &
\multicolumn{5}{c|}{\textbf{CQ}} &
\multicolumn{2}{c|}{\textbf{CO}} &
\multicolumn{2}{c|}{\textbf{CF}} &
\multicolumn{3}{c}{\textbf{TQ}} \\

\textbf{Retriever} & \textbf{Model} & \textbf{Method} &
\textbf{Prec.} & \textbf{Rec.} & \textbf{Soft Rec.} & \textbf{CSC} & \textbf{TRT} &
\textbf{CD} & \textbf{CDB} &
\textbf{PSS} & \textbf{CFA} &
\textbf{R-L} & \textbf{BS} & \textbf{LLM} \\

\midrule

% ===================== BM Sparse Retrieval =====================
\multicolumn{15}{l}{\textit{BM25 (Sparse Retrieval)}} \\

 & DeepSeek & RAG   & 0.105 & 0.133 & 0.838 & 0.512 & 0.766 & 0.864 & 0.425 & 0.541 & 0.257 & 0.116 & 0.551 & 0.090 \\
 & DeepSeek & ALCE   & 0.121 & 0.087 & 0.788 & 0.557 & 0.759 & 0.610 & 0.446 & 0.525 & 0.246 & 0.108 & 0.527 & 0.070 \\
 & DeepSeek & TAG   & 0.141 & 0.065 & 0.710 & 0.498 & 0.719 & 0.650 & 0.631 & 0.529 & 0.312 & 0.115 & 0.551 & 0.323 \\
 & DeepSeek & Workflow   & 0.105 & 0.127 & 0.835 & 0.526 & 0.773 & 0.856 & 0.380 & 0.537 & 0.241 & 0.107 & 0.537 & 0.025 \\
 & Llama & RAG   & 0.093 & 0.101 & 0.823 & 0.484 & 0.729 & 0.867 & 0.749 & 0.588 & 0.421 & 0.156 & 0.596 & 0.435 \\
 & Llama & ALCE   & 0.132 & 0.069 & 0.752 & 0.590 & 0.751 & 0.724 & 0.748 & 0.604 & 0.395 & 0.150 & 0.577 & 0.490 \\
 & Llama & TAG   & 0.137 & 0.053 & 0.724 & 0.567 & 0.759 & 0.622 & 0.839 & 0.569 & 0.362 & 0.152 & 0.573 & 0.347 \\
 & Llama & Workflow   & 0.090 & 0.083 & 0.768 & 0.496 & 0.689 & 0.828 & 0.763 & 0.597 & 0.413 & 0.148 & 0.589 & 0.347 \\
 & Qwen & RAG   & 0.096 & 0.101 & 0.813 & 0.528 & 0.774 & 0.780 & 0.568 & 0.608 & 0.394 & 0.150 & 0.590 & 0.455 \\
 & Qwen & ALCE   & 0.147 & 0.070 & 0.761 & 0.618 & 0.782 & 0.662 & 0.669 & 0.603 & 0.480 & 0.149 & 0.578 & 0.502 \\
 & Qwen & TAG   & 0.161 & 0.047 & 0.645 & 0.557 & 0.688 & 0.500 & 0.886 & 0.602 & 0.381 & 0.151 & 0.583 & 0.445 \\
 & Qwen & Workflow   & 0.095 & 0.063 & 0.739 & 0.517 & 0.719 & 0.728 & 0.700 & 0.590 & 0.400 & 0.148 & 0.599 & 0.427 \\
 & GPT-5.5 & RAG   & 0.089 & 0.140 & 0.855 & 0.534 & 0.781 & 0.930 & 0.643 & 0.607 & 0.350 & 0.123 & 0.581 & 0.713 \\
 & GPT-5.5 & ALCE   & 0.128 & 0.091 & 0.798 & 0.683 & 0.782 & 0.730 & 0.911 & 0.626 & 0.450 & 0.133 & 0.578 & 0.703 \\
 & GPT-5.5 & TAG   & 0.133 & 0.076 & 0.783 & 0.657 & 0.782 & 0.640 & 0.981 & 0.627 & 0.489 & 0.136 & 0.582 & 0.713 \\
 & GPT-5.5 & Workflow   & 0.113 & 0.161 & 0.865 & 0.566 & 0.781 & 0.894 & 0.710 & 0.605 & 0.386 & 0.120 & 0.583 & 0.695 \\

\midrule
% ===================== Dense Semantic Retrieval =====================
\multicolumn{15}{l}{\textit{Dense (Semantic Retrieval)}} \\

 & DeepSeek & RAG   & 0.139 & 0.150 & 0.818 & 0.490 & 0.721 & 0.864 & 0.512 & 0.564 & 0.283 & 0.118 & 0.557 & 0.163 \\
 & DeepSeek & ALCE   & 0.146 & 0.104 & 0.810 & 0.574 & 0.776 & 0.620 & 0.397 & 0.554 & 0.236 & 0.110 & 0.531 & 0.058 \\
 & DeepSeek & TAG   & 0.164 & 0.072 & 0.725 & 0.513 & 0.737 & 0.669 & 0.662 & 0.544 & 0.324 & 0.114 & 0.555 & 0.343 \\
 & DeepSeek & Workflow   & 0.123 & 0.150 & 0.837 & 0.511 & 0.759 & 0.877 & 0.449 & 0.546 & 0.258 & 0.110 & 0.544 & 0.095 \\
 & Llama & RAG   & 0.105 & 0.129 & 0.836 & 0.472 & 0.703 & 0.868 & 0.736 & 0.591 & 0.433 & 0.154 & 0.597 & 0.427 \\
 & Llama & ALCE   & 0.154 & 0.080 & 0.772 & 0.593 & 0.736 & 0.755 & 0.770 & 0.598 & 0.445 & 0.152 & 0.578 & 0.460 \\
 & Llama & TAG   & 0.160 & 0.064 & 0.736 & 0.576 & 0.746 & 0.634 & 0.810 & 0.574 & 0.389 & 0.154 & 0.577 & 0.362 \\
 & Llama & Workflow   & 0.103 & 0.108 & 0.773 & 0.487 & 0.674 & 0.824 & 0.757 & 0.589 & 0.413 & 0.147 & 0.583 & 0.372 \\
 & Qwen & RAG   & 0.132 & 0.113 & 0.811 & 0.532 & 0.769 & 0.834 & 0.578 & 0.608 & 0.402 & 0.152 & 0.592 & 0.502 \\
 & Qwen & ALCE   & 0.177 & 0.076 & 0.772 & 0.608 & 0.784 & 0.674 & 0.630 & 0.615 & 0.442 & 0.148 & 0.577 & 0.512 \\
 & Qwen & TAG   & 0.188 & 0.057 & 0.668 & 0.578 & 0.706 & 0.508 & 0.859 & 0.609 & 0.362 & 0.149 & 0.584 & 0.443 \\
 & Qwen & Workflow   & 0.102 & 0.070 & 0.759 & 0.551 & 0.750 & 0.762 & 0.679 & 0.606 & 0.422 & 0.147 & 0.592 & 0.458 \\
 & GPT-5.5 & RAG   & 0.104 & 0.175 & 0.879 & 0.528 & 0.783 & 0.986 & 0.699 & 0.605 & 0.363 & 0.123 & 0.580 & 0.662 \\
 & GPT-5.5 & ALCE   & 0.148 & 0.109 & 0.814 & 0.689 & 0.784 & 0.776 & 0.925 & 0.626 & 0.457 & 0.133 & 0.580 & 0.735 \\
 & GPT-5.5 & TAG   & 0.156 & 0.091 & 0.797 & 0.639 & 0.784 & 0.646 & 0.971 & 0.632 & 0.437 & 0.137 & 0.583 & 0.705 \\
 & GPT-5.5 & Workflow   & 0.115 & 0.178 & 0.869 & 0.559 & 0.783 & 0.940 & 0.719 & 0.613 & 0.393 & 0.119 & 0.581 & 0.682 \\

\midrule
% ===================== Oracle Ground Truth Citations =====================
\multicolumn{15}{l}{\textit{Oracle (Ground-Truth Citations)}} \\

 & DeepSeek & RAG   & 0.980 & 0.769 & 0.913 & 0.503 & 0.758 & 0.787 & 0.577 & 0.563 & 0.306 & 0.127 & 0.575 & 0.270 \\
 & DeepSeek & ALCE   & 0.980 & 0.719 & 0.902 & 0.583 & 0.766 & 0.628 & 0.485 & 0.530 & 0.253 & 0.115 & 0.549 & 0.142 \\
 & DeepSeek & TAG   & 0.950 & 0.543 & 0.826 & 0.506 & 0.742 & 0.715 & 0.661 & 0.545 & 0.316 & 0.124 & 0.565 & 0.360 \\
 & DeepSeek & Workflow   & 0.970 & 0.726 & 0.901 & 0.509 & 0.758 & 0.701 & 0.476 & 0.561 & 0.270 & 0.116 & 0.559 & 0.168 \\
 & Llama & RAG   & 0.990 & 0.785 & 0.935 & 0.489 & 0.712 & 0.853 & 0.718 & 0.600 & 0.410 & 0.163 & 0.609 & 0.485 \\
 & Llama & ALCE   & 0.980 & 0.542 & 0.856 & 0.632 & 0.759 & 0.749 & 0.800 & 0.609 & 0.442 & 0.156 & 0.586 & 0.490 \\
 & Llama & TAG   & 0.990 & 0.417 & 0.814 & 0.619 & 0.767 & 0.621 & 0.828 & 0.580 & 0.367 & 0.153 & 0.581 & 0.395 \\
 & Llama & Workflow   & 0.970 & 0.731 & 0.909 & 0.558 & 0.751 & 0.818 & 0.771 & 0.593 & 0.469 & 0.152 & 0.593 & 0.410 \\
 & Qwen & RAG   & 0.970 & 0.674 & 0.893 & 0.516 & 0.741 & 0.788 & 0.623 & 0.614 & 0.406 & 0.159 & 0.606 & 0.527 \\
 & Qwen & ALCE   & 1.000 & 0.470 & 0.849 & 0.619 & 0.782 & 0.684 & 0.729 & 0.607 & 0.465 & 0.152 & 0.587 & 0.555 \\
 & Qwen & TAG   & 0.840 & 0.278 & 0.661 & 0.527 & 0.656 & 0.510 & 0.919 & 0.609 & 0.380 & 0.152 & 0.590 & 0.465 \\
 & Qwen & Workflow   & 0.810 & 0.465 & 0.715 & 0.470 & 0.626 & 0.612 & 0.772 & 0.600 & 0.408 & 0.154 & 0.611 & 0.455 \\
 & GPT-5.5 & RAG   & 1.000 & 0.971 & 0.993 & 0.554 & 0.782 & 0.730 & 0.554 & 0.609 & 0.354 & 0.134 & 0.587 & 0.680 \\
 & GPT-5.5 & ALCE   & 1.000 & 0.768 & 0.941 & 0.665 & 0.782 & 0.774 & 0.984 & 0.628 & 0.463 & 0.140 & 0.584 & 0.710 \\
 & GPT-5.5 & TAG   & 1.000 & 0.652 & 0.904 & 0.605 & 0.782 & 0.714 & 0.968 & 0.625 & 0.479 & 0.140 & 0.590 & 0.708 \\
 & GPT-5.5 & Workflow   & 1.000 & 0.884 & 0.974 & 0.562 & 0.782 & 0.666 & 0.511 & 0.613 & 0.381 & 0.125 & 0.585 & 0.698 \\

\midrule
% ===================== Baselines =====================
\multicolumn{15}{l}{\textit{Baselines}} \\

 & DeepSeek & Direct   & 0.000 & 0.000 & 0.000 & -- & -- & 0.350 & 0.975 & 0.628 & 0.030 & 0.164 & 0.635 & 0.600 \\
 & Llama & Direct   & 0.000 & 0.000 & 0.000 & -- & -- & 0.350 & 0.958 & 0.588 & 0.030 & 0.173 & 0.639 & 0.490 \\
 & Qwen & Direct   & 0.000 & 0.000 & 0.000 & -- & -- & 0.350 & 0.952 & 0.603 & 0.030 & 0.169 & 0.641 & 0.540 \\
 & GPT-5.5 & Direct   & 0.000 & 0.000 & 0.000 & -- & -- & 0.350 & 1.000 & 0.603 & 0.064 & 0.127 & 0.573 & 0.720 \\

\midrule
 & Human & GT     & 1.000 & 1.000 & 1.000 & 0.866 & 0.902 & 0.838 & 0.941 & 1.000 & 1.000 & 1.000 & 1.000 & 0.525 \\

\bottomrule
\end{tabular}
}
\end{table*}

We use the main results to analyze where failures occur in scholarly positioning and which part of the pipeline produces them.
Table~\ref{tab:main_results} reports results across LLMs, representative generation paradigms, and multiple retrieval settings. The GPT-5.5 rows provide a frontier generator check while keeping the retrievers, candidate-list size, and citation mapping protocol identical to the other systems.
A retrieval-only diagnostic supports the same pattern. Among the top-50 RAG candidates, BM25 covers 0.144 of the matched gold citation set on average, while Dense covers 0.182. In the GPT-5.5 Workflow runs, multi-query retrieval covers 0.167 under BM25 and 0.186 under Dense. These values measure strict recovery of author-selected citations from a 1.09M-paper corpus, not missing corpus coverage, since all matched gold citation identifiers used for evaluation are present in the corpus.

\textbf{Finding 1: Abstract-level similarity is insufficient for citation selection decisions.}
Dense retrieval yields only modest gains in hard citation recall over BM25, even when using a semantic retriever. Under GPT-5.5 RAG, moving from BM25 to Dense improves hard recall from 0.140 to 0.175, but exact citation-set recovery remains low. The same pattern appears in Workflow, where Dense improves recall from 0.161 to 0.178. At the same time, soft citation recall is much higher than exact recall across retrieval-based systems. Many generated citations are therefore topically compatible alternatives even when they do not match the author citation set exactly. Topical compatibility, including TRT, is a useful diagnostic signal, but canonical citation selection requires finer-grained scholarly judgment about which prior works best position the target paper.

\textbf{Finding 2: Citation judgment failures are obscured by text-quality metrics.}
ROUGE-L, BERTScore, and LLM-as-Judge capture aspects of surface fluency and readability, but they do not track citation behavior. Direct prompting illustrates this decoupling: it receives competitive text-quality scores, including an LLM-as-Judge score of 0.720 for GPT-5.5 Direct, yet has zero citation recall because it has no supplied candidate list from which citation markers can be resolved. Models can therefore produce fluent, ordered prose while still failing the citation decisions that make related work academically useful.

\textbf{Finding 3: Models exhibit systematic deviations in organization and emphasis.}
Comparing generated sections with human-authored related work reveals consistent organizational deviations. Models often overconcentrate citations, as reflected by citation density and citation distribution balance, while PSS shows that their content blocks only partially follow the ordered positioning structure of author-written sections. GPT-5.5 improves writing quality and produces stronger PSS and CFA scores in several ALCE and TAG settings, but the scores remain far below the human reference. Stronger generation therefore improves local readability and some framing behavior without solving section-level organization.

\textbf{Finding 4: Correct citation selection does not guarantee appropriate citation placement and framing.}
Oracle experiments substantially increase hard and soft citation recall, confirming that retrieval is a major bottleneck. GPT-5.5 further shows how much a frontier generator can gain when the canonical references are supplied: Oracle-RAG reaches 0.971 hard recall and 0.993 soft recall, and Oracle-Workflow reaches 0.884 hard recall. However, their CSC, PSS, and CFA scores still remain well below the human reference. This shows that models struggle not only with \textit{which} papers to cite, but also \textit{where} and \textit{how} to use them as scholarly positioning moves. Citation placement, section-level positioning, and rhetorical framing therefore remain generation-level challenges beyond retrieval and topical relevance.

We also audit the citation-matching pipeline because small matching errors can distort low-recall citation metrics. Generated citation markers are mapped deterministically to the ordered candidate list supplied to the model. In Oracle runs, where all supplied candidates are matched gold citation identifiers, all 23,862 resolved citation IDs across 1,600 generations are members of the corresponding gold citation sets. Oracle precision below 1.0 therefore does not arise from post-hoc title matching noise. It is caused by 57 generations with no resolvable citation markers, which we report as empty-citation outputs and score as 0 in the main table; excluding these empty outputs gives oracle precision 1.000 over the remaining 1,543 generations. This reinforces the value of evaluating realized citation behavior rather than assuming that retrieved or provided citations are faithfully used.

Overall, the results show that related work generation involves several distinct decision layers with different failure modes. RWGBench makes these failures observable through positioning-oriented evaluation.

\subsection{Metric Complementarity and Case Study}

To examine whether RWGBench metrics provide distinct signals, we compute Spearman correlations across the 52 evaluated system configurations in Table~\ref{tab:metric_corr}. Hard and soft citation recall are correlated, as expected, but citation-set metrics are weakly correlated with PSS, CFA, and LLM-as-Judge. This supports the central premise that citation selection, citation support, structural organization, framing, and surface writing quality expose different parts of RWG behavior. PSS and LLM-as-Judge are strongly correlated in these experiments, plausibly because coherent section organization also improves judged readability. Neither captures citation-set recovery, which is why citation-level metrics remain necessary.

\begin{table}[t]
\centering
\caption{Spearman correlations among representative metrics across evaluated systems.}
\label{tab:metric_corr}
\setlength{\tabcolsep}{3pt}
\renewcommand{\arraystretch}{1.02}
\resizebox{\columnwidth}{!}{%
\begin{tabular}{lrrrrrrr}
\toprule
 & Rec. & Soft & CSC & TRT & PSS & CFA & LLM \\
\midrule
Rec. & 1.00 & 0.89 & -0.13 & 0.16 & 0.05 & 0.16 & 0.02 \\
Soft & 0.89 & 1.00 & 0.00 & 0.40 & 0.08 & 0.19 & 0.08 \\
CSC & -0.13 & 0.00 & 1.00 & 0.68 & 0.55 & 0.42 & 0.53 \\
TRT & 0.16 & 0.40 & 0.68 & 1.00 & 0.47 & 0.19 & 0.52 \\
PSS & 0.05 & 0.08 & 0.55 & 0.47 & 1.00 & 0.54 & 0.88 \\
CFA & 0.16 & 0.19 & 0.42 & 0.19 & 0.54 & 1.00 & 0.45 \\
LLM & 0.02 & 0.08 & 0.53 & 0.52 & 0.88 & 0.45 & 1.00 \\
\bottomrule
\end{tabular}
}
\end{table}

Table~\ref{tab:case_study} provides representative qualitative checks. These cases illustrate the diagnostic role of the metrics. Soft recall identifies plausible alternatives when exact recall is low, Oracle recall isolates generation-level framing failures, and text-quality metrics can remain high even when citations are ungrounded.

\begin{table*}[t]
\centering
\caption{Representative diagnostic cases. Metrics highlight distinct failure modes that are difficult to see from surface text quality alone.}
\label{tab:case_study}
\setlength{\tabcolsep}{4pt}
\renewcommand{\arraystretch}{1.05}
\resizebox{\textwidth}{!}{%
\begin{tabular}{p{0.26\textwidth}p{0.15\textwidth}p{0.17\textwidth}p{0.34\textwidth}}
\toprule
\textbf{Target paper} & \textbf{System} & \textbf{Key scores} & \textbf{Diagnostic observation} \\
\midrule
DRESSing Up LLM & GPT-5.5 RAG Dense & Rec.=0.000, Soft Rec.=0.984 & The generated citations are topically compatible with the target area but do not match the author citation set, illustrating why exact recall and soft recall should be reported separately. \\
DN-4DGS & GPT-5.5 Oracle-RAG & Rec.=1.000, CFA=0.000 & The model cites the canonical papers but fails to reproduce comparable citation-frame usage, showing that citation-set recovery does not solve scholarly framing. \\
Communicating Activations Between Language Model Agents & GPT-5.5 Direct & LLM=1.000, Rec.=0.000 & The section receives the highest writing-quality score despite having no grounded citation recall, illustrating the decoupling between fluent prose and citation decision quality. \\
\bottomrule
\end{tabular}
}
\end{table*}

\subsection{Evaluation Consistency Analysis}

To assess the robustness of the LLM-as-Judge text-quality evaluation, we additionally run Claude Opus and Gemini Pro with the same rubric on a sampled subset. Because the metric is used primarily for system comparison, we focus on decisive pairwise preferences where both judges assign different scores. The audit covers 265 scored sections in total. It includes the original 200-section consistency sample and 65 GPT-5.5 outputs sampled uniformly across GPT-5.5 configurations. On non-tied item pairs, GPT-5.5 agrees with Claude and Gemini on 95.3\% and 93.4\% of pairwise orderings, respectively. These comparisons cover 62.1\% and 55.1\% of all item pairs. The corresponding Goodman and Kruskal's $\gamma$ values are 0.906 and 0.869. The GPT-only subset receives saturated external scores, which limits within-GPT ranking, but adding these outputs does not weaken cross-judge ordering consistency. These results support the use of LLM-as-Judge as an auxiliary writing-quality metric. Since GPT-5.5 is also included as a generator, our claims about frontier generation rely on citation, organization, and framing metrics computed independently of the LLM judge.

\subsection{Human Validation}

We further conduct a blinded human validation study to examine whether the proposed metrics reflect human preferences along their intended dimensions. Six graduate annotators completed the same 60 tasks, yielding 360 task submissions and 1,260 item-level responses across the multiple-choice dimensions. Completion logs show that all annotators finished all tasks, with no missing tasks or duplicate submissions. For section-level A/B judgments, tie votes are retained in the logs and treated as non-decisive for directional agreement. We aggregate preferences by unique majority and report counts for every analysis set. As an inter-annotator consistency check, a position-balance sensitivity analysis yields unique majorities for 28 to 30 of 30 section-pair tasks across the annotation dimensions. The average winning-vote share ranges from 70.4\% to 74.0\%. For metric alignment, we compare automatic metrics with the human-majority direction on decisive comparisons. High-consensus cases require at least 75\% of non-tie human votes. Clear-margin cases are defined as the upper third of non-tied metric margins. The study serves as supplementary validation of directional metric behavior, complementing the full automatic benchmark rather than replacing it.

\begin{table*}[t]
\centering
\caption{Human consistency and metric-human alignment in the validation study. Counts are shown for transparency.}
\label{tab:human_validation}
\setlength{\tabcolsep}{4pt}
\renewcommand{\arraystretch}{1.08}
\begin{tabularx}{\textwidth}{p{0.18\textwidth}p{0.18\textwidth}Xp{0.16\textwidth}}
\toprule
\textbf{Validation target} & \textbf{Signal} & \textbf{Protocol} & \textbf{Agreement} \\
\midrule
Human consistency & Section preferences & position-balance sensitivity & 28 to 30/30; 70.4 to 74.0\% \\
Writing quality & LLM-as-Judge & high-consensus preference & 5/5; 100.0\% \\
Scholarly positioning & PSS & non-direct high-consensus & 8/11; 72.7\% \\
Scholarly positioning & PSS & non-direct clear-margin & 6/7; 85.7\% \\
Overall usefulness & Soft Recall & position-balance sensitivity, non-direct & 13/17; 76.5\% \\
Overall usefulness & Soft Recall & position-balance sensitivity, clear-margin & 5/6; 83.3\% \\
Overall usefulness & CSC & non-direct clear-margin & 7/7; 100.0\% \\
\bottomrule
\end{tabularx}
\end{table*}

Table~\ref{tab:human_validation} shows that alignment is strongest when each metric is evaluated on its intended preference dimension. LLM-as-Judge agrees with all high-consensus human writing-quality preferences, and with 11/17 decisive writing-quality preferences overall. PSS aligns with human judgments of scholarly positioning, especially outside direct prompting comparisons and when the PSS margin is clear. Soft Recall aligns with overall usefulness under the position-balance sensitivity analysis, and CSC matches all seven clear support-quality comparisons. Overall, the human study provides supplementary evidence that RWGBench's diagnostic metrics capture distinct aspects of related work quality.

\section{Related Work}
\label{sec:related}
\subsection{Automated Scientific Writing}
Recent advances in large language models have enabled progress in long-form scientific text generation. Prior work shows that scientific writing requires maintaining discourse coherence and argumentative structure beyond sentence-level fluency~\cite{krishna2021hurdles,gehrmann2022repairing}.
Survey generation systems~\cite{autosurvey,surge,gao2024llm} synthesize broad overviews of research areas, emphasizing topical coverage. In contrast, related work generation is a \textit{paper-centric} task that positions a specific target paper within existing literature. Genre and discourse studies highlight that citation choice and organization in related work serve rhetorical functions rather than content summarization~\cite{ZHANG2022101076,PETRIC2007238}.
Recent RWG approaches incorporate target-awareness~\cite{tag}, and survey-style reviews summarize existing methods~\cite{li2024related}. However, they consistently note the absence of standardized benchmarks and evaluation protocols for document-level scholarly positioning.

\subsection{Citation Recommendation and Generation}
Citation recommendation systems identify relevant papers based on textual similarity, structural signals, or citation context~\cite{bhagavatula2018content,wright2021citeworth,celik2025citebart}. Complementary work on citation sentence generation focuses on producing locally appropriate citation statements~\cite{xing2020automatic,wright2024citeassist}.
Additional studies analyze citation intent and discourse roles~\cite{cohan2019structural,PETRIC2007238}, while grounded generation frameworks such as ALCE~\cite{alce} encourage explicit evidence attribution.
Although these efforts provide important components for RWG systems, they mainly operate at the level of individual citations or sentences, evaluating \textit{local correctness} rather than coherent citation selection, organization, and framing at the section level.

\subsection{Retrieval-Augmented Generation}
Retrieval-Augmented Generation (RAG) enhances large language models by conditioning generation on documents retrieved from an external corpus~\cite{lewis2020retrieval,robertson2009probabilistic}. Extensions such as active, graph-based, and agentic RAG broaden the retrieval and reasoning process~\cite{jiang2023active,edge2024local,jin2025search}. Although RAG improves grounding and knowledge access, its evaluation typically focuses on retrieval accuracy, factual consistency, or answer quality. In scholarly writing, retrieval alone is insufficient: cited papers must be strategically integrated into argumentative structures, and their roles must align with the rhetorical goals of the related work section. RWGBench targets this underexplored gap by evaluating how retrieved papers are selected, placed, and framed in document-level related work generation.

\subsection{Benchmarks for Scientific Text Generation}

Evaluation of scientific text generation commonly relies on lexical overlap~\cite{lin2004rouge,papineni2002bleu}, semantic similarity~\cite{zhang2019bertscore}, factuality metrics~\cite{min2023factscore}, and LLM-based judgments~\cite{zheng2023judging,liu2023geval}. While effective for surface-level assessment, these metrics provide limited insight into citation decision-making.
Recent benchmarks have begun to address broader evaluation needs. OARelatedWork~\cite{docekal2024oarelatedwork} provides a large-scale dataset of 94,450 papers with full-text cited documents. Its evaluation relies on ROUGE and BERTScore and does not address citation decision-making. RWGBench instead frames RWG as a citation decision-making task, providing citation-specific metrics and a large retrieval corpus for end-to-end evaluation under realistic conditions.
GREP~\cite{grep} proposes an LLM-as-judge evaluation framework with fine-grained dimensions and contrastive examples. Our evaluation results (Section~\ref{sec:exp}) suggest that general writing-quality judgments should not replace targeted citation behavior metrics when evaluating scholarly positioning.
DeepScholar-Bench~\cite{deepscholar2025} evaluates generative research synthesis across retrieval quality, synthesis accuracy, and verifiability, reporting aggregate scores that offer limited diagnosis of citation-level failures.
RWGBench adopts a complementary philosophy of \textit{behavioral diagnosis}: fine-grained metrics that isolate failure modes in citation selection, placement, and organization, combined with Oracle experiments that disentangle retrieval- and generation-level limitations.

\section{Conclusion}

We presented RWGBench, a benchmark designed to evaluate related work generation from the perspective of scholarly positioning and citation decision-making. 
RWGBench consists of 100 carefully curated peer-reviewed computer science papers with author-written related work sections, supported by a retrieval corpus of over 1.09 million papers and a multi-dimensional evaluation framework. Experimental results demonstrate that current retrieval-based systems struggle with citation selection, and GPT-5.5 runs show that stronger generators improve writing quality and citation recovery under matched retrieval conditions. When canonical references are provided, GPT-5.5 substantially improves citation recall, yet generation-level challenges, particularly citation placement, organization, and framing, persist beyond citation-set recovery.
By providing reproducible infrastructure and interpretable evaluation signals, RWGBench aims to support systematic, targeted progress in automated scholarly writing.

\bibliographystyle{ACM-Reference-Format}
\bibliography{references}

\end{document}